%Paper: hep-th/9403080
%From: "J.W.van de Leur" <vdleur@math.ruu.nl>
%Date: Tue, 15 Mar 1994 14:03:32 +0100
%Date (revised): Tue, 17 May 1994 14:40:01 +0200

%%%%%%%%%%%%%%%%%%%%%%%%%%%%%%%%
%%%                          %%%
%%%  Use amstex version 2.1  %%%
%%%                          %%%
%%%%%%%%%%%%%%%%%%%%%%%%%%%%%%%%

\documentstyle{amsppt}

\NoBlackBoxes
\magnification=\magstep1
\hsize=6.5truein
\vsize=8.5truein
\document
\baselineskip=.15truein

\topmatter
\title   KdV Type Hierarchies, the String Equation and $W_{1+\infty}$
Constraints
\endtitle
\author
Johan van de Leur
\endauthor
\thanks
\;The author was partly supported by a fellowship of the Royal Netherlands
Academy of
Arts and Sciences. E-mail: vdleur\@math.ruu.nl
\endthanks
\address
J.W. van de Leur, Kievitdwarsstraat 22,  3514VE Utrecht, The Netherlands
\endaddress
\abstract
To every partition $n=n_1+n_2+\cdots+n_s$ one can associate a vertex
operator realization of the Lie algebras $a_{\infty}$ and $\hat{gl}_n$.
Using this construction  we make reductions of the $s$--component KP hierarchy,
reductions which are related to these partitions.
In this way we obtain matrix KdV type
equations. Now assuming that (1) $\tau$ is a $\tau$--function of the
$[n_1,n_2,\ldots,n_s]$--th reduced KP hierarchy and
(2) $\tau$  satisfies a `natural' string equation,
we prove that $\tau$ also satisfies the vacuum constraints of the
$W_{1+\infty}$ algebra.
\endabstract
\endtopmatter

\subheading{\S 0. Introduction}
\vskip 10pt

In recent years KdV type hierarchies have been related to 2D gravity. To be
slightly more precise  (see [Dij] for the details and references),
the square root of the partition function of the Hermitian $(n-1)$--matrix
model
in the continuum limit is the $\tau$--function of the $n$--reduced
Kadomtsev Petviashvili (KP) hierarchy. Hence, the $(n-1)$--matrix  model
corresponds to $n$--th Gelfand Dickey hierarchy. For $n=2,3$ these hierarchies
are better known as the KdV-- and Boussinesque hierarchy, respectively.
The partition function is then characterized by the so-called string
equation:
$$L_{-1}\tau=\frac{1}{n}\frac{\partial \tau}{\partial x_1},\tag{0.1}$$
where $L_{-1}$ is an element of  the $c=n$  Virasoro algebra,
wich is related to the principal realization of the affine lie algebra
$\hat{sl}_n$,
or rather $\hat{gl}_n$. Let $\alpha_k=-kx_{-k},\ 0,\ \frac{\partial}{\partial
x_k}$ for
$k<0,\ k=0,\ k>0$, respectively, then
$$L_k=\frac{1}{2n}\sum_{\ell\in {\Bbb Z}}:\alpha_{-\ell}\alpha_{\ell+nk}:
+\delta_{0k}\frac{n^2-1}{24n}.
\tag{0.2}
$$

By making the shift $x_{n+1}\mapsto x_{n+1}+{n\over n+1}$, we modify the
origin of the $\tau$--function and thus obtain the following
form of the string equation:
$$L_{-1}\tau=0.\tag{0.3}$$

Actually, it can be shown ([FKN] and [G]) that the above conditions,
$n$--th reduced KP and equation (0.3) (which from now on we will call
the string equation),
on a $\tau$--function of the KP hierarchy imply
more general constraints, viz. the vacuum constraints of the $W_{1+\infty}$
algebra. This last condition is reduced to the vacuum conditions of the $W_n$
algebra when some redundant variables are eliminated.

The $W_{1+\infty}$ algebra is the central extension of the Lie algebra of
differential
operators on ${\Bbb C}^{\times}$.
This central extension was discovered by Kac and Peterson in 1981 [KP3] (see
also [Ra], [KRa]). It has as basis the operators
$W^{(\ell+1)}_k=-t^{k+\ell}(\frac{\partial}{\partial t})^{\ell}$, $\ell\in{\Bbb
Z}_+$,
$k\in{\Bbb Z}$, together with the central element $c$. There is a well-known
way
how to express these elements in the elements of the Heisenberg algebra,
the $\alpha_k$'s. The $W_{1+\infty}$ constraints then are
$$\hat W^{(\ell+1)}_k\tau=\{W^{(\ell+1)}_k +\delta_{k,0}c_{\ell}\}\tau=0
\ \text{for } -k\le \ell\ge 0.
\tag{0.4}
$$
For the above $\tau$--function, $\hat W^{(1)}_k=-\alpha_{nk}$ and
$\hat W^{(2)}_k=L_k-\frac{nk+1}{n}\alpha_{nk}$.

It is well--known that the $n$--reduced KP hierarchy is related to the
principal realization (a vertex realization) of the basic module of
$\hat{sl}_n$. However there are many inequivalent vertex realization.
Kac and Peterson [KP1] and independently Lepowsky [L] showed that
for the basic representation of a simply--laced affine Lie algebra
these different realizations are parametrized by the conjugacy classes
of the Weyl group of the corresponding finite dimensional Lie algebra.
Hence, for the case of $\hat{sl}_n$ they are parametrized by the partitions
$n=n_1+n_2+\cdots+n_s$ of $n$. An explicit description of these realizations
was given in [TV] (see also $\S 2$). There the construction was given in such a
way
that it was possible to make reductions of the KP--hierarchy.
In all these constructions a `natural' Virasoro algebra played an important
role.
A natural question now is: If $\tau$ is a $\tau$--function of this
$[n_1,n_2,\ldots,n_s]$--th
reduced KP hierarchy and $\tau$ satisfies the string equation (0.3),
where $L_{-1}$ is an element of this new Virasoro algebra, does $\tau$ also
satisfy some
corresponding $W_{1+\infty}$ constraints? In this paper we give a positive
answer to this question. As will be shown in $\S 6$, there exists a`natural'
$W_{1+\infty}$  algebra for which (0.4) holds.

This paper is organized as follows. Sections 1--3 give results wich
were obtained in [KV] and [TV] (see also [BT]). Its major part is an
exposition of the $s$--component KP hierarchy following [KV].
In $\S1$, we describe the semi--infinite
wedge representation of the group $GL_{\infty}$ and the Lie algebras
$gl_{\infty}$ and $a_{\infty}$. We  define the KP hierarchy
in the so--called fermionic picture. The loop algebra $\hat{gl}_n$ is
introduced in
$\S 2$. We obtain it as a subalgebra of $a_{\infty}$.
Next we construct to every partition  $n=n_1+n_2+\cdots+n_s$ of $n$ a vertex
operator
realization of $a_\infty$ and $\hat{gl}_n$. $\S 3$ is devoted to the
description of $s$--component
KP hierarchy in terms of formal pseudo--differential operators. $\S 4$
describes
reductions of this $s$--component KP hierarchy related to the above
partitions.
In $\S 5$ we introduce the string equation and deduce its consequences in
terms of the pseudo--differential operators. Using the results of $\S 5$ we
deduce
in $\S 6$ the $W_{1+\infty}$ constraints. $\S 7$ is devoted to a geometric
interpretation of the string equation  on the Sato Grassmannian,
which is similar to that of [KS].

Notice that since the Toda lattice hierarchy of [UT] is related to the
2--component
KP hierarchy, some results of this paper also hold for certain reductions
of the Toda lattice hierarchy.

Finally, I would like to thank Victor Kac for valuable discussions, and
the Mathematical Institute of the University of Utrecht
for the computer and e-mail facilities. I want to dedicate this paper to the
memory of my father, Noud van de Leur, who died quite suddenly in the period
that I was writing this article.

\vskip 10pt
\subheading{\S 1.  The semi-infinite wedge
representation of the group $GL_{\infty}$ and the KP hierarchy in
the fermionic picture}

\vskip 10pt
{\bf 1.1.}  Consider the infinite complex matrix group
$$GL_{\infty} = \{ A = (a_{ij})_{i,j \in {\Bbb Z}+\frac{1}{2}}|A\
\text{is invertible and all but a finite number of}\ a_{ij} -
\delta_{ij}\ \text{are}\ 0\}$$
and its Lie algebra
$$gl_{\infty} = \{ a = (a_{ij})_{i,j \in {\Bbb Z}+\frac{1}{2}}|\
\text{all but a finite number of}\ a_{ij}\ \text{are}\ 0\}$$
with bracket $[a,b] = ab-ba$.  This Lie algebra has a
basis consisting of matrices $E_{ij},\ i,j \in {\Bbb Z} + \frac{1}{2}$, where
$E_{ij}$ is the matrix with a $1$ on the $(i,j)$-th entry and zeros
elsewhere. Now $gl_{\infty}$ is a subalgebra of the bigger Lie algebra
$$\overline{gl_{\infty}} = \{ a = (a_{ij})_{i,j \in {\Bbb Z}+\frac{1}{2}}
|\  a_{ij}=0\  \text{if}\   |i-j|>>0\}.$$
This Lie algebra $\overline{gl_{\infty}}$ has a universal central extension
$a_{\infty} := \overline{gl_{\infty}}\bigoplus {\Bbb C}c$
with Lie bracketdefined by
$$[a+\alpha c,b+\beta c]=ab-ba+\mu (a,b)c , \tag{1.1.1}$$
for $a,b\in \overline{gl_{\infty}}$ and $\alpha ,\beta\in{\Bbb C}$;
here $\mu$ is the following 2--cocycle:
$$\mu (E_{ij},E_{kl})=\delta_{il}\delta_{jk}(\theta(i)-\theta(j)),
\tag{1.1.2}$$
where the linear mapping $\theta :{\Bbb R}\to {\Bbb C}$ is defined by
$$\theta(i):=\cases 0 & \text{if}\ i>0,\\
		    1 & \text{if}\ i\le 0.\endcases \tag{1.1.3}$$
Let ${\Bbb C}^{\infty} = \bigoplus_{j \in {\Bbb Z}+\frac{1}{2}} {\Bbb
C} v_{j}$ be an infinite dimensional complex vector space with fixed
basis $\{ v_{j}\}_{j \in {\Bbb Z}+\frac{1}{2}}$.  Both the group
$GL_{\infty}$ and the Lie algebras $gl_{\infty}$ and $a_{\infty}$ act linearly
on
${\Bbb C}^{\infty}$ via the usual formula:
$$E_{ij} (v_{k}) = \delta_{jk} v_{i}.$$
We introduce, following [KP2], the corresponding semi-infinite wedge space $F =
\Lambda^{\frac{1}{2}\infty} {\Bbb C}^{\infty}$, this is the vector space
with a basis consisting of all semi-infinite monomials of the form
$v_{i_{1}} \wedge v_{i_{2}} \wedge v_{i_{3}} \ldots$, where $i_{1} >
i_{2} > i_{3} > \ldots$ and $i_{\ell +1} = i_{\ell} -1$ for $\ell >>
0$.  We can now define representations $R$ of $GL_{\infty}$ on $F$ by
$$
R(A) (v_{i_{1}} \wedge v_{i_{2}} \wedge v_{i_{3}} \wedge \cdots) = A
v_{i_{1}} \wedge Av_{i_{2}} \wedge Av_{i_{3}} \wedge \cdots . \tag{1.1.4}
$$
In order to describe representations of the Lie algebras we find it convenient
to define  wedging and contracting
operators $\psi^{-}_{j}$ and $\psi^{+}_{j}\ \ (j \in {\Bbb Z} +
\frac{1}{2})$ on $F$ by
$$\align
&\psi^{-}_{j} (v_{i_{1}} \wedge v_{i_{2}} \wedge \cdots ) = \cases 0
& \text{if}\ j = i_{s} \text{for some}\ s \\
(-1)^{s} v_{i_{1}} \wedge v_{i_{2}} \cdots \wedge v_{i_{s}} \wedge
v_{-j} \wedge v_{i_{s+1}} \wedge \cdots &\text{if}\ i_{s} > -j >
i_{s+1}\endcases \\
&\psi^{+}_{j} (v_{i_{1}} \wedge v_{i_{2}} \wedge \cdots ) = \cases 0
&\text{if}\ j \neq i_{s}\ \text{for all}\ s \\
(-1)^{s+1} v_{i_{1}} \wedge v_{i_{2}} \wedge \cdots \wedge
v_{i_{s-1}} \wedge v_{i_{s+1}} \wedge \cdots &\text{if}\ j = i_{s}.
\endcases
\endalign
$$
Notice that the definition of $\psi^{\pm}_{j}$ differs from the one in [KV].
The reason for this will become cclear in $\S$ 7 where we describe the
connection with the Sato Grassmannian.
These  wedging and contracting operators satisfy the following relations
$(i,j \in {\Bbb Z}+\frac{1}{2}, \lambda ,\mu = +,-)$:
$$\psi^{\lambda}_{i} \psi^{\mu}_{j} + \psi^{\mu}_{j}
\psi^{\lambda}_{i} = \delta_{\lambda ,-\mu} \delta_{i,-j}, \tag{1.1.5}$$
hence they generate a Clifford algebra, which we denote by ${\Cal C}\ell$.

Introduce the following elements of $F$ $(m \in {\Bbb Z})$:
$$|m\rangle = v_{m-\frac{1}{2} } \wedge v_{m-\frac{3}{2} } \wedge
v_{m-\frac{5}{2} } \wedge \cdots .$$
It is clear that $F$ is an irreducible ${\Cal C}\ell$-module such that
$$\psi^{\pm}_{j} |0\rangle = 0 \ \text{for}\ j > 0 . \tag{1.1.6}$$
We are now able to define representations $r$, $\hat r$ of $gl_{\infty}$,
$a_{\infty}$ on $F$ by
$$r(E_{ij})=\psi^-_{-i}\psi^+_j,\quad \hat
r(E_{ij})=:\psi^-_{-i}\psi^+_j:,\quad \hat r(c)=I,
$$
where $:\ :$ stands for the {\it normal ordered product} defined in
the usual way $(\lambda ,\mu = +$ or $-$):
$$:\psi^{\lambda (i)}_{k} \psi^{\mu (j)}_{\ell}: = \cases \psi^{
\lambda (i)}_{k}
\psi^{\mu (j)}_{\ell}\ &\text{if}\ \ell \ge k \\
-\psi^{\mu (j)}_{\ell} \psi^{\lambda (i)}_{k} &\text{if}\ \ell <
k.\endcases \tag{1.1.7}$$
\vskip 10pt
{\bf 1.2.} Define the {\it charge decomposition}
$$F = \bigoplus_{m \in {\Bbb Z}} F^{(m)} \tag{1.2.1}$$
by letting

$$\text{charge}(|0\rangle ) = 0\ \text{and charge} (\psi^{\pm}_{j}) =
\pm 1. \tag{1.2.2}$$
It  is easy to see that each $F^{(m)}$ is irreducible with
respect to $g\ell_{\infty}$, $a_{\infty}$ (and $GL_{\infty}$).  Note that
$|m\rangle$ is its highest weight vector, i.e.,
$\hat r(E_{ij})=r(E_{ij})-\delta_{ij}\theta(i)$ and
$$\align
&r(E_{ij})|m\rangle = 0 \ \text{for}\ i < j, \\
&r(E_{ii})|m\rangle = 0\  (\text{resp.}\ = |m\rangle ) \ \text{if}\ i > m\
(\text{resp. if}\ i < m).
\endalign
$$
Let
${\Cal O}
= R(GL_{\infty})|0\rangle \subset F^{(0)}$
be the $GL_{\infty}\text{-orbit}$
of the vacuum vector $|0\rangle$, then one has

\proclaim{Proposition 1.1 ([KP2])}  A non-zero element $\tau$ of $F^{(0)}$
lies in ${\Cal O}$ if and only if the following equation holds in $F \otimes
F$:
$$\sum_{k \in {\Bbb Z}+\frac{1}{2}} \psi^{+}_{k} \tau \otimes \psi^{-}_{-k}
\tau = 0. \tag{1.2.3}$$
\endproclaim

\demo{Proof}For a proof  see [KP2] or [KR].\ \ \ $\square$
\enddemo

Equation (1.2.3) is called the {\it KP hierarchy in the fermionic
picture}.

\vskip 10pt
\subheading{\S 2.  The loop algebra $\hat{gl}_n$, partitions of $n$ and vertex
operator constructions}

\vskip 10pt
{\bf 2.1}  Let $\tilde{gl}_n =gl_n({\Bbb C}[t,t^{-1}])$ be the loop algebra
associated to $gl_n({\Bbb C})$. This algebra has a natural representation
on the vector space $({\Bbb C}[t,t^{-1}])^n$. Let $\{ w_i\}$
 be the standard basis of ${\Bbb C}^n$, by identifying  $({\Bbb
 C}[t,t^{-1}])^n$ over ${\Bbb C}$ with ${\Bbb C}^{\infty}$ via
$v_{nk+j-\frac{1}{2}}=t^{-k}w_j$
we obtain an embedding $\phi :\tilde{gl}_n \to \overline{gl_{\infty}}$;
$$\phi (t^k e_{ij})=\sum_{\ell\in {\Bbb Z}} E_{n(\ell -k)+i-\frac{1}{2},
n\ell+j-\frac{1}{2}},$$
here $e_{ij}$ is a basis of $gl_n({\Bbb C})$.

A straightforward calculation shows that the restriction of the cocycle $\mu$
to $\phi (\tilde {gl}_n)$ induces the following 2--cocycle
 on
$\tilde{gl}_n$:
$$\mu (x(t),y(t))= \text{Res}_{t=0} dt\ \text{tr} ({dx(t)\over dt}y(t)).$$
Here and further $\text{Res}_{t=0} dt \sum_j f_jt^j$ stands for $f_{-1}$. This
gives a central extension $\hat{gl}_n=\tilde{gl}_n\bigoplus {\Bbb C}K$, where
the bracket is defined by
$$[t^\ell x+\alpha K,t^my+\beta K]=t^{\ell+m}(xy-yx)+\ell
\delta_{\ell,-m}\text{tr}(xy)K.$$
In this way we have an embedding $\phi :\hat{gl}_n\to a_{\infty}$, where
$\phi(K)=c$.

Since $F$ is a module for $a_{\infty}$, it is clear that with this
embedding we also have a representation of $\hat{gl}_n$
on this semi-infinite wedge space. It is well--known that the level one
representations of  the affine Kac--Moody algebra
$\hat{gl}_n$ have a lot of inequivalent realizations. To be more precise, Kac
and Peterson [KP1] and independently Lepowsky [L] showed
that to every conjugacy class of the Weyl group of $gl_n({\Bbb C})$
or rather $sl_n({\Bbb C})$ there exists an inequivalent vertex operator
realization of the same level one module. Hence to every partition
of $n$, there exists such a construction.

We will now sketch how one can  construct these vertex realizations of
$\hat{gl}_n$,
following [TV]. From now on let $n=n_1+ n_2+\cdots+n_s$ be a partition of $n$
into $s$ parts,  and denote by
$N_a=n_1+n_2+\cdots +n_{a-1}$.
We begin by relabeling the basis vectors $v_j$ and with them the corresponding
fermionic (wedging and contracting) operators:
($1\le a \le s,\ 1\le p\le n_a,\ j\in {\Bbb Z}$)
$$\align
v^{(a)}_{n_aj-p+\frac{1}{2}}&=v_{nj-N_a-p+\frac{1}{2}} \\
\psi^{\pm (a)}_{n_a \mp p\pm \frac{1}{2}}&=\psi^{\pm }_{n_aj\mp N_a\mp
p\pm\frac{1}{2}}
\tag{2.1.1}
\endalign$$
Notice that with this relabeling we have:
$\psi^{\pm (a)}_{k}|0\rangle = 0\ \text{for}\ k > 0.$
We also rewrite the $E_{ij}$'s:
$$E^{(ab)}_{n_aj-p+\frac{1}{2},n_bk-q+\frac{1}{2}}=E_{nj-N_a-p+\frac{1}{2},nk-N_b-q+\frac{1}{2}}.$$
The corresponding Lie bracket on $a_{\infty}$ is given by
$$[E^{(ab)}_{jk},E^{(cd)}_{\ell m}]=\delta_{bc}\delta_{kl}E^{(ad)}_{jm}-
\delta_{ad}\delta_{jm}E^{(db)}_{\ell
k}+\delta_{ad}\delta_{bc}\delta_{jm}\delta_{k\ell}(\theta (j)-\theta (k))c,$$
and $\hat r(E^{(ab)}_{jk})=:\psi^{-(a)}_{-j}\psi^{+b}_k:$.

Introduce the  fermionic fields $(z \in {\Bbb C}^{\times})$:
$$\psi^{\pm (a)}(z) \overset{\text{def}}\to{=} \sum_{k \in {\Bbb
Z}+\frac{1}{2}} \psi^{\pm
(a)}_{k} z^{-k-\frac{1}{2}}.\tag{2.1.2}$$
Let $N$ be the least common multiple of $n_1,n_2,\ldots,n_s$.
It was shown in [TV] that the modes of the fields
$$:\psi^{+(a)}(\omega_a^p z^{N\over n_a})\psi^{-(b)}(\omega_b^q z^{N\over
n_b}):, \tag{2.1.3}$$
for $1\le a,b\le s$, $1\le p\le n_a$, $1\le q\le n_b$, where $\omega_a=e^{2\pi
i/n_a}$, together with the identity,
generate a representation of $\hat{gl}_n$ with  $K=1$.

Next we introduce special bosonic fields ($1\le a\le s$):
$$\alpha^{(a)}(z) \equiv \sum_{k \in {\Bbb Z}} \alpha^{(a)}_{k} z^{-k-1}
\overset{def}\to{=} :\psi^{+(a)}(z) \psi^{-(a)}(z):. \tag{2.1.4}$$
The operators $\alpha^{(a)}_{k} $
satsify the canonical commutation relation of the associative
oscillator algebra,  which we
denote by ${\frak a}$:
$$[\alpha^{(i)}_{k},\alpha^{(j)}_{\ell}] =
k\delta_{ij}\delta_{k,-\ell},\tag{2.1.5}$$
and one has
$$\alpha^{(i)}_{k}|m\rangle = 0 \ \text{for}\ k > 0.\tag{2.1.6}$$
It is easy to see that restricted to $\hat{gl}_n$,
$F^{(0)}$ is its basic highest weight representation (see [K, Chapter
12]).

In order to express the fermionic fields $\psi^{\pm (i)}(z)$ in terms of
the bosonic fields $\alpha^{(i)}(z)$, we need some additional operators
$Q_{i},\ i = 1,\ldots ,s$, on $F$.  These operators are uniquely defined by
the following conditions:
$$Q_{i}|0\rangle = \psi^{+(i)}_{-\frac{1}{2}} |0\rangle ,\ Q_{i}\psi^{\pm
(j)}_{k} = (-1)^{\delta_{ij}+1} \psi^{\pm
(j)}_{k\mp \delta_{ij}}Q_{i}.\tag{2.1.7}$$
They satisfy the following commutation relations:
$$Q_{i}Q_{j} = -Q_{j}Q_{i}\ \text{if}\ i \neq j,\ [\alpha^{(i)}_{k},Q_{j}] =
\delta_{ij} \delta_{k0}Q_{j}.\tag{2.1.8}$$

\proclaim{Theorem 2.1}  ([DJKM1], [JM])
$$\psi^{\pm (i)}(z) = Q^{\pm 1}_{i}z^{\pm \alpha^{(i)}_{0}} \exp
(\mp \sum_{k < 0} \frac{1}{k} \alpha^{(i)}_{k}z^{-k})\exp(\mp
\sum_{k > 0} \frac{1}{k} \alpha^{(i)}_{k} z^{-k}). \tag{2.1.9}$$
\endproclaim

\demo{Proof} See [TV].
\enddemo

The operators on the right-hand side of (2.1.9) are called vertex
operators.  They made their first appearance in string theory (cf.
[FK]).

If one substitutes (2.1.9) into (2.1.3), one obtains the vertex operator
realization of $\hat{gl}_n$ which is related to the partition
$n=n_1+n_2+\cdots+n_s$ (see [TV] for more details).

\vskip 10pt
{\bf 2.2.} The realization  of $\hat{gl}_n$, described in the previous section,
has a natural Virasoro algebra. In [TV], it was shown that the following two
sets of operators have the same action on $F$.
$$\align
&L_k=\sum_{i=1}^s\{ \sum_{j\in{\Bbb Z}}{1\over
2n_i}:\alpha_{-j}^{(i)}\alpha_{j+n_ik}^{(i)}:+\delta_{k0}{{n_i^2-1}\over
24n_i}\},\tag{2.2.1} \\
&H_k=\sum_{i=1}^s\{\sum_{j\in{\Bbb Z}+\frac{1}{2}}({j\over n_i}+{k\over
2}):\psi^{+(i)}_{-j}\psi^{-(i)}_{j+n_ik}:+\delta_{k0}{{n_i^2-1}\over
24n_i}\}, \tag{2.2.2} \endalign
$$
So $L_k=H_k$,
$$[L_k,\psi_j^{\pm (i)}]=-(\frac{j}{n_i}+\frac{k}{2})\psi_{j+n_ik}^{\pm(i)}
\tag{2.2.3}
$$
 and
$$[L_k,L_\ell ]=(k-\ell)L_{k+\ell}+\delta_{k,-\ell}{{k^3-k}\over 12}n. $$

\vskip 10pt

{\bf 2.3.} We will now use the results of \S 2.1 to
describe  the $s$-component boson-fermion
correspondence.  Let ${\Bbb C}[x]$ be the space of polynomials in
indeterminates $x = \{ x^{(i)}_{k}\},\ k = 1,2,\ldots ,\ i =
1,2,\ldots ,s$.  Let $L$ be a lattice with a basis $\delta_{1},\ldots
,\delta_{s}$ over ${\Bbb Z}$ and the symmetric bilinear form
$(\delta_{i}|\delta_{j}) = \delta_{ij}$, where $\delta_{ij}$ is the
Kronecker symbol.  Let
$$\varepsilon_{ij} = \cases -1 &\text{if $i > j$} \\
1 &\text{if $i \leq j$.} \endcases \tag{2.3.1}$$
Define a bimultiplicative function $\varepsilon :\ L \times L @>>> \{
\pm 1 \}$ by letting
$$\varepsilon (\delta_{i}, \delta_{j}) = \varepsilon_{ij}.
\tag{2.3.2}$$
Let $\delta = \delta_{1} + \ldots + \delta_{s},\  Q= \{ \gamma \in
L|\ (\delta | \gamma ) = 0\}$, $\Delta = \{ \alpha_{ij} :=
\delta_{i}-\delta_{j}| i,j = 1,\ldots ,s,\ i \neq j \}$.  Of course
$Q$ is the root lattice of $sl_{s}({\Bbb C})$, the set $\Delta$
being the root system.

Consider the vector space ${\Bbb C}[L]$ with basis $e^{\gamma}$,\
$\gamma \in L$, and the following twisted group algebra product:
$$e^{\alpha}e^{\beta} = \varepsilon (\alpha ,\beta)e^{\alpha +
\beta}. \tag{2.3.3}$$
Let $B = {\Bbb C}[x] \otimes_{\Bbb C} {\Bbb C}[L]$ be the tensor
product of algebras.  Then the $s$-component boson-fermion
correspondence is the vector space isomorphism
$$\sigma :F @>\sim >> B, \tag{2.3.4}$$
given by
$$\sigma (\alpha^{(i_{1})}_{-m_{1}} \ldots
\alpha^{(i_{r})}_{-m_{r}}Q_1^{k_{1}}\ldots Q_s^{k_{s}}|0 \rangle ) = m_{1}
\ldots
m_{s}x^{(i_{1})}_{m_{1}} \ldots x^{(i_{r})}_{m_{r}} \otimes
e^{k_{1}\delta_{1} + \ldots + k_{s}\delta_{s}} . \tag{2.3.5}$$
The transported charge then will be as follows:
$$
\text{charge}(p(x)\otimes e^{\gamma}) = (\delta |\gamma).
\tag{2.3.6}
$$
We denote the transported charge decomposition by
$$B = \bigoplus_{m \in {\Bbb Z}} B^{(m)}.$$

The transported action of the operators $\alpha^{(i)}_{m}$ and $Q_{j}$ looks
as follows:
$$\cases
\sigma \alpha^{(j)}_{-m}\sigma^{-1}(p(x) \otimes e^{\gamma}) =
mx^{(j)}_{m}p(x)\otimes e^{\gamma},\ \text{if}\ m > 0, &\  \\
\sigma \alpha^{(j)}_{m} \sigma^{-1}(p(x) \otimes e^{\gamma}) = \frac{\partial
p(x)}{\partial x_{m}} \otimes e^{\gamma},\ \text{if}\ m > 0, &\  \\
\sigma \alpha^{(j)}_{0} \sigma^{-1} (p(x) \otimes e^{\gamma}) =
(\delta_{j}|\gamma ) p(x) \otimes e^{\gamma} , &\ \\
\sigma Q_{j} \sigma^{-1} (p(x) \otimes e^{\gamma}) = \varepsilon
(\delta_{j},\gamma)  p(x) \otimes e^{\gamma + \delta_{j}}
. & \
\endcases \tag{2.3.7}
$$
For notational conveniences, we introduce
$\delta_j=\sigma\alpha_0^{(j)}\sigma^{-1}$.
Notice that $e^{\delta_j}=\sigma Q_j\sigma^{-1}$.
\vskip 10pt
{\bf 2.4.}  Using the isomorphism $\sigma$ we can reformulate the KP hierarchy
(1.2.3) in the bosonic picture.

We start by observing that (1.2.3) can be rewritten as follows:
$$\text{Res}_{z=0}\ dz ( \sum^{s}_{j=1} \psi^{+(j)}(z)\tau
\otimes \psi^{-(j)}(z)\tau ) = 0,\ \tau \in F^{(0)}.
\tag{2.4.1}$$
Notice that for $\tau \in F^{(0)},\ \sigma (\tau) = \sum_{\gamma \in Q}
\tau_{\gamma}(x)e^{\gamma}$.
 Here and further  we write $\tau_{\gamma}(x)e^{\gamma}$ for
$\tau_{\gamma} \otimes
e^{\gamma}$.  Using Theorem 2.1, equation (2.4.1) turns under $\sigma
\otimes \sigma :\ F \otimes F \overset\,\,\sim\to\longrightarrow
{\Bbb C}[x^{\prime},x^{\prime \prime}]
\otimes ({\Bbb C}[L^{\prime}] \otimes {\Bbb C}[L^{\prime \prime}])$ into the
following set of equations;
 for all $\alpha ,\beta \in L$ such that $(\alpha
|\delta ) = -(\beta |\delta ) = 1$ we have:
$$\aligned
&\text{Res}_{z=0} ( dz
 \sum^{s}_{j=1} \varepsilon (\delta_{j}, \alpha-\beta)
z^{(\delta_{j}|\alpha - \beta - 2\delta_{j})}  \\
 &\times \exp
(\sum^{\infty}_{k=1} (x^{(j)^{\prime}}_{k} - x^{(j)^{\prime
\prime}}_{k})z^{k})
\exp (-\sum^{\infty}_{k=1} (\frac{\partial}{\partial x^{(j)^{\prime}}_{k}}
 - \frac {\partial}{\partial x^{(j)^{\prime
\prime}}_{k}})\frac{z^{-k}}{k}) \\
& \tau_{\alpha -
\delta_{j}}(x^{\prime})(e^{\alpha})^{\prime}\tau_{\beta + \delta_{j}}(x^{\prime
\prime})(e^{\beta})^{\prime\prime})
= 0  . \endaligned \tag{2.4.2}$$

\vskip 10pt
\subheading{\S 3. The algebra of formal pseudo-differential operators and the
$s$-component KP hierarchy as a dynamical system}

\vskip 10pt
{\bf 3.0.} The KP hierarchy and its $s$-component generalizations admit
several  formulations.  The one we will give here was  introduced by Sato [S],
it is given in the
language of formal pseudo-differential operators.  We will show that this
formulation follows from the $\tau$-function formulation given by
equation (2.4.2).

\vskip 10pt
{\bf 3.1.} We shall work over the algebra ${\Cal A}$ of formal power
series over ${\Bbb C}$ in indeterminates $x = (x^{(j)}_{k})$, where
$k = 1,2,\ldots $ and $j = 1,\ldots ,s$.  The indeterminates
$x^{(1)}_{1},\ldots ,x^{(s)}_{1}$ will be viewed as variables and
$x^{(j)}_{k}$ with $k \geq 2$ as parameters.  Let
$$\partial = \frac{\partial}{\partial x^{(1)}_{1}} + \ldots +
\frac{\partial}{\partial x^{(s)}_{1}}.$$
{\it A formal} $s \times s$ {\it matrix pseudo-differential
operator} is an expression of the form
$$P(x,\partial ) = \sum_{j \leq N} P_{j}(x)\partial^{j}, \tag{3.1.1}$$
where $P_{j}$ are $s \times s$ matrices over ${\Cal A}$.   Let $\Psi$ denote
the vector
space over ${\Bbb C}$ of all expressions (3.1.1).  We have a linear
isomorphism $S :\ \Psi @>>> \text{Mat}_{s}({\Cal A}((z)))$ given by
$S(P(x,\partial )) = P(x,z)$.  The matrix series
$P(x,z)$ in indeterminates $x$ and $z$ is called the {\it symbol} of
$P(x,\partial )$.

Now we may define a product $\circ$ on $\Psi$ making it an associative
algebra:
$$S(P \circ Q) = \sum^{\infty}_{n = 0} \frac{1}{n!}
\frac{\partial^{n}S(P)}{\partial z^{n}} \partial^{n}S(Q).\tag{3.1.2}$$
{}From now on, we shall drop the multiplication sign $\circ$ when no ambiguity
may arise.
One defines the differential part of $P(x,\partial )$ by
$P_{+}(x,\partial ) = \sum^{N}_{j=0} P_{j}(x) \partial^{j},$
and let $P_{-} = P-P_{+}$.  We have the corresponding vector space
decomposition:
$$\Psi = \Psi_{-} \oplus \Psi_{+}. \tag{3.1.3}$$
One defines a linear map $*: \Psi \rightarrow \Psi$ by the
following formula:
$$(\sum_{j} P_{j}\partial^{j})^{*} = \sum_{j} (-\partial )^{j}
\circ ^{t}\! P_{j}. \tag{3.1.4}$$
Here and further $^{t}P$ stands for the transpose of the matrix $P$.
Note that $*$ is an anti-involution of the algebra
$\Psi$.

\vskip 10pt
{\bf 3.2.}  Introduce the following notation
$$z \cdot x^{(j)} = \sum^{\infty}_{k=1} x^{(j)}_{k} z^{k},\ e^{z
\cdot x} = diag (e^{z\cdot x^{(1)}} ,\ldots ,e^{z\cdot x^{(s)}} ).$$
The algebra $\Psi$ acts on the space $U_{+}$ (resp.
$U_{-}$) of formal oscillating matrix functions of the form
$$\sum_{j \leq N} P_{j}z^{j}e^{z\cdot x}\ \ \ (\text{resp.}\ \sum_{j
\leq N} P_{j}z^{j} e^{-z\cdot x}), \ \text{where}\ \ P_{j} \in \
\text{Mat}_{s} ( {\Cal A} ),$$
in the obvious way:
$$P(x)\partial^{j}e^{\pm z\cdot x} = P(x)(\pm z)^{j} e^{\pm z\cdot x}.$$
One has the following fundamental lemma (see [KV]).

\proclaim{Lemma 3.1} If $P,Q \in \Psi$ are such that
$$Res_{z=0} (P(x,\partial) e^{z\cdot x})  \ ^{t}
(Q(x^{\prime},\partial^{\prime}) e^{-z\cdot x^{\prime}})  dz = 0,
\tag{3.2.1}$$
then $(P \circ Q^{*})_{-} = 0$.
\endproclaim

\vskip 10pt
{\bf 3.3.}  We proceed now to rewrite the formulation (2.4.2) of the
$s$-component KP hierarchy in terms of formal pseudo-differential
operators.

For each $\alpha \in \ \text{supp}\ \tau :=\{\alpha\in Q| \tau=\sum_{\alpha\in
Q}\tau_{\alpha}e^{\alpha}, \tau_{\alpha}\ne 0\}$ we define the (matrix
valued) functions
$$V^{\pm} (\alpha ,x,z) = (V^{\pm}_{ij}(\alpha ,x,z))^{s}_{i,j=1}
\tag{3.3.1}$$
as follows:
$$\aligned
&V^{\pm}_{ij}(\alpha ,x,z) \overset{\text{def}}\to{=}
\varepsilon (\delta_{j} , \alpha + \delta_{i})
 z^{(\delta_{j}|\pm \alpha + \alpha_{ij})} \\
& \times \exp (\pm \sum^{\infty}_{k=1} x^{(j)}_{k} z^{k})
\exp(\mp \sum^{\infty}_{k=1} \frac{\partial}{\partial
x^{(j)}_{k}} \frac{z^{-k}}{k}) \tau_{\alpha  \pm
\alpha_{ij}}  (x)/\tau_{\alpha}(x) .
\endaligned \tag{3.3.2}
$$
It is easy to see that equation (2.4.2) is equivalent to the
following bilinear identity:
$$Res_{z=0}V^{+}(\alpha ,x,z)\ ^{t}V^{-}(\beta ,x^{\prime},z)dz = 0\
\text{for all}\ \alpha ,\beta \in Q. \tag{3.3.3}$$
Define $s \times s$ matrices $W^{\pm (m)} (\alpha ,x)$ by the
following generating series (cf. (3.3.2)):
$$
\sum^{\infty}_{m=0}
W^{\pm (m)}_{ij} (\alpha ,x)(\pm z)^{-m}
= \varepsilon_{ji}z^{\delta_{ij}-1} (\exp \mp
\sum^{\infty}_{k=1} \frac{\partial}{\partial x^{(j)}_{k}}\frac{z^{-k}}{k})
\tau_{\alpha \pm
\alpha_{ij}} (x))/\tau_{\alpha} (x). \tag{3.3.4}
$$

We see from (3.3.2) that $V^{\pm}(\alpha ,x,z)$ can be written in the
following form:
$$V^{\pm}(\alpha ,x,z) = (\sum^{\infty}_{m=0}
W^{\pm (m)}(\alpha ,x)R^{\pm}(\alpha ,\pm z)(\pm
z)^{-m})e^{\pm z \cdot x}, \tag{3.3.5}$$
where
$$R^{\pm}(\alpha ,z) = \sum^{s}_{i=1}
\varepsilon (\delta_{i}, \alpha ) E_{ii} (\pm z)^{\pm
(\delta_{i}|\alpha )}. \tag{3.3.6}$$
Here and further $E_{ij}$ stands for the $s \times s$ matrix whose
$(i,j)$ entry is $1$ and all other entries are zero.  Now it is clear
that $V^{\pm}(\alpha ,x,z)$ can be
written in
terms of formal pseudo-differential operators
$$P^{\pm}(\alpha ) \equiv P^{\pm} (\alpha ,x,\partial ) =
I_{n} + \sum^{\infty}_{m=1} W^{\pm (m)} (\alpha ,x)\partial^{-m}\
\text{and}\ R^{\pm}(\alpha ) = R^{\pm}(\alpha ,\partial) \tag{3.3.7}$$
as follows:
$$V^{\pm}(\alpha ,x,z) = P^{\pm } (\alpha )
R^{\pm}(\alpha)e^{\pm z \cdot x} . \tag{3.3.8}$$
Since obviously
$R^{-}(\alpha ,\partial)^{-1} = R^{+}(\alpha
,\partial)^{*}$,
using Lemma 3.1 we deduce from the bilinear identity
(3.3.3):
$$\align
&P^{-}(\alpha  ) = (P^{+}(\alpha  )^{*})^{-1} ,
\tag{3.3.9}\\
&(P^{+}(\alpha)R^{+}(\alpha - \beta)P^{+}(\beta)^{-1})_{-} = 0\
\text{for all}\ \alpha ,\beta \in \text{supp}\ \tau . \tag{3.3.10}
\endalign$$

Victor Kac and the author showed in [KV] that
given $\beta \in \text{supp}\ \tau$, all the
pseudo-differential operators $P^+(\alpha )$, $\alpha \in \text{supp}\
\tau$, are
completely determined by $P^+(\beta )$ from equations (3.3.10). They also
showed that $P=P^+(\alpha)$ satisfies the Sato equation:
$$
\frac{\partial P}{\partial x^{(j)}_{k}} = -(PE_{jj} \circ
\partial^{k} \circ P^{-1})_{-} \circ P. \tag{3.3.11}$$
To be more precise, one has the following

\proclaim{Proposition 3.2}  Consider the formal oscillating functions
$V^{+}(\alpha ,x,z)$ and $V^{-}(\alpha ,x,z),$ \newline $\alpha \in Q$, of the
form (3.3.8), where $R^{\pm}(\alpha ,z)$ are given by (3.3.6) and
$P^{\pm}(\alpha ,x,\partial ) \in I_{s} + \Psi_{-}$.  Then the
bilinear identity (3.3.3) for all $\alpha ,\beta \in \ \text{supp}\
\tau$ is equivalent to the Sato equation (3.3.11) for each $P =
P^{+}(\alpha )$ and the matching conditions (3.3.9-10)  for
all $\alpha ,\beta \in \ \text{supp} \ \tau$.
\endproclaim

\vskip 10pt
{\bf 3.4.}  Fix $\alpha \in Q$, introduce the following formal
pseudo-differential
operators $L(\alpha),\  C^{(j)}(\alpha )$, and
differential operators  $B^{(j)}_{m}(\alpha)$:
$$\aligned
L \equiv L(\alpha )
  & = P^{+}(\alpha) \circ \partial \circ P^{+}(\alpha)^{-1}, \\
C^{(j)} \equiv C^{(j)}(\alpha ) &=
P^{+}(\alpha)E_{jj} P^{+}(\alpha)^{-1}, \\
B^{(j)}_{m} \equiv B^{(j)}_{m}(\alpha) &=
(P^{+}(\alpha)E_{jj} \circ \partial^{m} \circ
P^{+}(\alpha)^{-1})_{+}.
\endaligned \tag{3.4.1}
$$
Then
$$\aligned
L &= I_{s} \partial + \sum^{\infty}_{j=1} U^{(j)}(x) \partial^{-j} , \\
C^{(i)} &= E_{ii} + \sum^{\infty}_{j=1} C^{(i,j)}(x)
\partial^{-j},\ i = 1,2,\cdots ,s,
\endaligned \tag{3.4.2}$$
subject to the conditions
$$
\sum^{s}_{i=1} C^{(i)} = I_{s}, \
C^{(i)}L = LC^{(i)},\ C^{(i)}C^{(j)} = \delta_{ij} C^{(i)}.
 \tag{3.4.3}$$
They satisfy the following set of equations for some $P \in I_{n} + \Psi_{-}$:
$$
\cases LP = P\partial &\ \\
C^{(i)}P = PE_{ii} &\ \\
\displaystyle{\frac{\partial P}{\partial x^{(i)}_{k}} = -(L^{(i)k})_{-} P,\
\text{where}\  L^{(i)} = C^{(i)}L.}
\endcases
\tag{3.4.4}$$

\proclaim{Proposition 3.3}  The system of equations (3.4.4) has a
solution $P \in I_{s} + \Psi_{-}$ if and only if we can find a formal
oscillating function of the form
$$W(x,z) = (I_{s} + \sum^{\infty}_{j=1} W^{(j)}(x)z^{-j})e^{z\cdot x}
\tag{3.4.5}$$
that satisfies the linear equations
$$
LW = zW ,\
C^{(i)}W = WE_{ii} , \
\frac{\partial W}{\partial x^{(i)}_{k}} = B^{(i)}_{k}W .
 \tag{3.4.6}$$
\endproclaim
And finally, one has the following
\proclaim{Proposition 3.4} If for every $\alpha \in Q$ the
formal pseudo-differential operators $L \equiv L(\alpha)$ and $C^{(j)} \equiv
C^{(j)}(\alpha)$ of the form (3.4.2) satisfy conditions (3.4.3) and if the
equations (3.4.4) have a solution $P \equiv P(\alpha) \in I_{s} +
\Psi_{-}$, then the
differential operators $B^{(j)}_{k} \equiv B^{(j)}_{k}(\alpha)$
satisfy one of the following equivalent
conditions:
$$\left\{ \matrix \displaystyle{\frac{\partial L}{\partial x^{(j)}_{k}} =
[B^{(j)}_{k},L],}  & \   \\
\displaystyle{\frac{\partial C^{(i)}}{\partial x^{(j)}_{k}} =
[B^{(j)}_{k},C^{(i)}],} & \
\  \endmatrix \right. \tag{3.4.7}$$
$$\frac{\partial L^{(i)}}{\partial x^{(j)}_{k}} =
[B^{(j)}_{k},L^{(i)}], \tag{3.4.8}$$
$$\frac{\partial B^{(i)}_{\ell}}{\partial x^{(j)}_{k}} -
\frac{\partial B^{(j)}_{k}}{\partial x^{(i)}_{\ell}} =
[B^{(j)}_{k},B^{(i)}_{\ell}]. \tag{3.4.9}$$
Here $L^{(j)}\equiv L^{(j)}(\alpha)=C^{(j)}(\alpha)\circ L(\alpha)$.
\endproclaim
Equations (3.4.7) and (3.4.8) are called {\it Lax type} equations.
Equations (3.4.9) are called the {\it Zakharov-Shabat type}
equations.  The latter
are the compatibility conditions for the linear problem (3.4.6).

\vskip 10pt
\subheading{\S4 . $[n_1,n_2,\ldots ,n_s]$-reductions of the $s$-component KP
hierarchy}

\vskip 10pt
{\bf 4.1.} Using (2.1.9), (2.1.3), (2.3.5) and (2.3.7),
we obtain the to the partition
$n=n_1+n_2+\cdots +n_s$ related vertex operator realization of
$\hat{gl}_n$ in the vector space $B^{(m)}$.
Now, restricted to $\hat{sl}_n$, the representation in $F^{(m)}$ is not
irreducible anymore, since $\hat{sl}_n$ commutes with the operators
$$\beta_{kn_s}^{(s)}=\sqrt{n_s\over N}\sum_{i=1}^s \alpha_{kn_i}^{(i)},\quad
k\in {\Bbb Z}.
\tag{4.1.1}$$
In order to describe the irreducible part of the representation of $\hat{sl}_n$
in $B^{(0)}$ containing the vacuum vector 1, we choose the complementary
generators of the oscillator algebra $\frak a$
contained in $\hat{sl}_n$ ($k\in {\Bbb Z}$):
$$\beta^{(j)}_{k} = \cases \alpha^{(j)}_{k} &\text{if $k \notin
n_j{\Bbb Z}$,} \\
{{N_{j+1}\alpha_{\ell n_{j+1}}^{(j+1)}-n_{j+1}(\alpha_{\ell n_1}^{(1)}
+\alpha_{\ell n_2}^{(2)}+\cdots +\alpha_{\ell n_j}^{(j)})}\over
\sqrt{N_{j+1}(N_{j+1}-n_{j+1})}}&\text{if $k=\ell n_j$ and
$1 \leq j < s$,}
\endcases \tag{4.1.2}$$
so that the operators (4.1.1 and 2) also satisfy relations (2.1.5). Hence,
introducing the new indeterminates
$$y^{(j)}_{k} = \cases x^{(j)}_{k} &\text{if $k \notin n_j{\Bbb N}$}, \\
{{N_{j+1}x_{\ell n_{j+1}}^{(j+1)}-(n_1x_{\ell n_1}^{(1)}+n_2x_{\ell
n_2}^{(2)}+\cdots +n_jx_{\ell n_j}^{(j)})}\over
\sqrt{N_{j+1}(N_{j+1}-n_{j+1})}}&\text{if $k=\ell n_j$ and $1 \leq j < s$,}\\
{{n_1x_{\ell n_1}^{(1)}+n_2x_{\ell n_2}^{(2)}+\cdots +n_sx_{\ell
n_s}^{(s)}}\over \sqrt{Nn_s}} &\text{if $k=\ell n_s$ and $j=s$},
\endcases \tag{4.1.3} $$
we have: ${\Bbb C}[x] = {\Bbb C}[y]$ and
$$\sigma (\beta^{(j)}_{k}) = \frac{\partial}{\partial y^{(j)}_{k}}\
\text{and}\ \sigma (\beta^{(j)}_{-k}) = ky^{(j)}_{k} \ \text{if}\ k
> 0. \tag{4.1.4}$$
Now it is clear that the irreducible with respect to
$\hat{sl}_n$ subspace of $B^{(0)}$ containing the
vacuum $1$ is the vector space
$$B^{(0)}_{[n_1,n_2,\ldots ,n_s]} = {\Bbb C}[y^{(j)}_{k}|1 \leq j < s,\ k \in
{\Bbb
N},\ \text{or}\ j = s,\ k \in {\Bbb N}\backslash n_s{\Bbb Z}] \otimes
{\Bbb C}[Q]. \tag{4.1.5}$$
The vertex operator realization of $\hat{sl}_n$ in
the vector space $B^{(0)}_{[n_1,n_2,\ldots ,n_s]}$ is then obtained
by expressing the
fields (2.1.3) in terms of vertex
operators (2.1.9), which are expressed via (4.1.2) in
the operators (4.1.4), the
operators $e^{\delta_i -\delta_j}$ and $\delta_i-\delta_j\
(1 \leq i < j \leq s)$ (see [TV] for details).

The $s$-component KP hierarchy of  equations  (2.4.2) on $\tau
\in B^{(0)} = {\Bbb C}[y] \otimes {\Bbb C}[Q]$ when restricted to
$\tau \in B^{(0)}_{[n_1,n_2,\ldots ,n_s]}$ is called the {\it $[n_1,n_2,\ldots
,n_s]$-th reduced} KP {\it
hierarchy}.  It is obtained from the $s$-component KP hierarchy by
making the change of variables (4.1.3) and putting zero all terms
containing partial derivates by
$y^{(s)}_{n_s},y^{(s)}_{2n_s},y^{(s)}_{3n_s},\ldots$.

The totality of solutions of the $[n_1,n_2,\ldots ,n_s]$-th reduced KP
hierarchy is given
by the following

\proclaim{Proposition 4.1}  Let ${\Cal O}_{[n_1,n_2,\ldots ,n_s]}$ be the orbit
of $1$
under the (projective) representation of the loop group
$SL_{n}({\Bbb C}[t,t^{-1}])$ corresponding to the representation of
$\hat{sl}_n$ in $B^{(0)}_{[n_1,n_2,\ldots ,n_s]}$.  Then
$${\Cal O}_{[n_1,n_2,\ldots ,n_s]} = \sigma ({\Cal O}) \cap
B^{(0)}_{[n_1,n_2,\ldots ,n_s]}.$$
In other words, the $\tau$-functions of the $[n_1,n_2,\ldots ,n_s]$-th reduced
KP hierarchy are
precisely the $\tau$-functions of the KP hierarchy in the variables
$y^{(j)}_{k}$, which are independent of the variables
$y^{(s)}_{\ell n_s},\ \ell \in {\Bbb N}$.
\endproclaim

\demo\nofrills{Proof} \ is the same as of a similar statement in [KP2].
\ \ \ $\square$
\enddemo

\vskip 10pt
{\bf 4.2.} It is clear from the definitions and results of \S 4.1 that the
condition on the $s$-component KP hierarchy to be $[n_1,n_2,\ldots ,n_s]$-th
reduced
is equivalent to
$$\sum^{s}_{j=1} \frac{\partial \tau}{\partial x^{(j)}_{kn_j}}
= 0,\quad\text{for all $k\in {\Bbb N}$}\tag{4.2.1}$$
Using the Sato equation (3.3.11), this
implies the
following  two equivalent conditions:
$$\sum^{s}_{j=1} \frac{\partial W(\alpha )}{\partial x^{(j)}_{kn_j}} =
W(\alpha)\sum_{j=1}^sz^{kn_j}E_{jj}, \tag{4.2.2}$$
$$(\sum_{j=1}^sL(\alpha)^{kn_j}C^{(j)})_{-}=0.\tag{4.2.3}$$

\vskip 10pt
\subheading{\S 5. The string equation}

\vskip 10pt
{\bf 5.1.} {\it From now on we assume that $\tau$ is any solution of the KP
hierarchy.} In particular, we no longer assume that $\tau_{\alpha}$ is a
polynomial. For instance the soliton and dromion solutions of [KV, \S 5] are
allowed. Of course this means that the corresponding wave functions
$V^{\pm}(\alpha ,z)$ will be of a more general nature than before.

Recall from \S 3 the wave function $V(\alpha,z)\equiv
V^+(\alpha,z)=P(\alpha)R(\alpha)e^{z\cdot x}=$\break
$
P^+(\alpha)R^+(\alpha)e^{z\cdot x}$. It is natural to compute
$$\align
{{\partial V(\alpha,z)}\over \partial z}
&={\partial\over\partial z}P(\alpha)R(\alpha)e^{z\cdot x}\\
&=P(\alpha)R(\alpha){\partial\over\partial z}e^{z\cdot x}\\
&=P(\alpha)R(\alpha)\sum_{a=1}^s\sum_{k=1}^{\infty} kx_k^{(a)}
\partial^{k-1}E_{aa}R(\alpha)^{-1}P(\alpha)^{-1}V(\alpha,z).
\endalign$$
Define
$$M(\alpha):=P(\alpha)R(\alpha)\sum_{a=1}^s\sum_{k=1}^{\infty} kx_k^{(a)}
\partial^{k-1}E_{aa}R(\alpha)^{-1}P(\alpha)^{-1},\tag{5.1.1}$$
then one easily checks that $[L(\alpha),M(\alpha)]=1$ and
$$[\sum_{a=1}^s L(\alpha)^{n_a}C^{(a)}(\alpha), M(\alpha)\sum_{a=1}^s {1\over
n_a}L(\alpha)^{1-n_a}C^{(a)}(\alpha)]=1.\tag{5.1.2}$$

Next, we calculate the $(i,j)$--th coefficient of $(M(\alpha)\sum_{a} {1\over
n_a}L(\alpha)^{1-n_a}C^{(a)}(\alpha))_{-}P(\alpha)R(\alpha)$.
Let $P=S(P(\alpha))$ and $R=S(R(\alpha))$, then
$$\align
&S((M(\alpha)\sum_{a=1}^s{1 \over n_a}L(\alpha)^{1-n_a}C^{(a)}(\alpha))_-
P(\alpha)R(\alpha))_{ij})= \\
&S(P(\alpha)R(\alpha)\sum_{a=1}^s\sum_{k\in{\Bbb N}}{1\over n_a
}kx_k^{(a)}\partial^{k-n_a}
E_{aa}R(\alpha)^{-1}P(\alpha)^{-1})_-P(\alpha)R(\alpha))_{ij})= \\
&({{\partial PR}\over\partial z}\sum_{a=1}^s{1\over n_a}E_{aa}z^{1-n_a}+
\sum_{a=1}^s\{
 \sum_{k=1}^{n_a}{k\over n_a}x_k^{(a)}PRE_{aa}z^{k-n_a}
  -x^{(a)}_{n_a}E_{aa}PR \\
  &\qquad -\sum_{k=1}^{\infty} {{k+n_a}\over n_a}x^{(a)}_{k+n_a}{{\partial
  PR}\over \partial x_k^{(a)}}
\})_{ij}
\endalign$$
Define
$$\align
\tilde\tau_{\alpha+\delta_i-\delta_j}&=\exp(-\sum_{k=1}^{\infty}
{\partial\over\partial x_k^{(j)}}{{z^{-k}}\over
k})\tau_{\alpha+\delta_i-\delta_j}
(x)\\
&=\tau_{\alpha+\delta_i-\delta_j}(\ldots,x_k^{(b)}-{{\delta_{jb}}\over
kz^k},\ldots),
\endalign
$$
then
$$(PR)_{ij}=\epsilon(\delta_j |\alpha+\delta_i)z^{\delta_{ij}-1+(\delta_j
|\alpha)}
{\tilde\tau_{\alpha+\delta_i-\delta_j}\over \tau_{\alpha}}
$$
and hence
$$
\align
&S((M(\alpha)\sum_{a=1}^s{1\over
n_a}L(\alpha)^{1-n_a}C^{(a)}(\alpha))_-P(\alpha)R(\alpha))_{ij})=
{{\epsilon(\delta_j|\alpha+\delta_i)}\over n_j}\times \\
&\{ {1\over\tau_{\alpha}}(\sum_{k=1}^{\infty}{\partial\over\partial x_k^{(j)}}
z^{-k-n_j}+(\delta_{ij}-1+(\delta_j|\alpha))z^{-n_j})\tilde\tau_{\alpha+\delta_i-\delta_j}
+\sum_{k=1}^{n_j} kx_k^{(j)}{{\tilde{\tau}_{\alpha+\delta_i-\delta_j}}\over
\tau_{\alpha}}z^{k-n_j} \\
&-n_jx^{(i)}_{n_i}{{\tilde{\tau}_{\alpha+\delta_i-\delta_j}}\over\tau_{\alpha}}
-\sum_{a=1}^s {{n_j}\over n_a}\sum_{k=1}^{\infty} (k+n_a)x_{k+n_a}^{(a)}
{\partial\over \partial
x_k^{(a)}}({{\tilde{\tau}_{\alpha+\delta_i-\delta_j}}\over \tau_{\alpha}})
\} z^{\delta_{ij}-1+(\delta_j|\alpha)}.
\tag{5.1.3}
\endalign
$$
\vskip 10pt
{\bf 5.2.} We introduce the natural generalization of the string equation
(0.3).
Let $L_{-1}$ be given by (2.2.1), the string equation is the following
constraint
on $\tau\in F^{(0)}$:
$$L_{-1}\tau=0. \tag{5.2.1}$$
Using (2.3.7) we rewrite $L_{-1}$ in terms of operators on $B^{(0)}$:
$$L_{-1}=\sum_{a=1}^s \{
\delta_a x_{n_a}^{(a)}+
{1\over 2n_a}\sum_{p=1}^{n_a-1} p(n_a-p)x_p^{(a)}x_{n_a-p}^{(a)}
+{1\over n_a}\sum_{k=1}^{\infty}(k+n_a)x_{k+n_a}^{(a)}{\partial\over\partial
x_k^{(a)}}\}.
$$
Since $\tau=\sum_{\alpha\in Q}\tau_{\alpha}e^{\alpha}$ and $L_{-1}\tau=0$,
we find that for all $\alpha\in Q$:
$$\sum_{a=1}^s \{
(\delta_a|\alpha)x_{n_a}^{(a)}+
{1\over 2n_a}\sum_{p=1}^{n_a-1} p(n_a-p)x_p^{(a)}x_{n_a-p}^{(a)}
+{1\over n_a}\sum_{k=1}^{\infty}(k+n_a)x_{k+n_a}^{(a)}{\partial\over\partial
x_k^{(a)}}\}
\tau_{\alpha}=0.
\tag{5.2.2}
$$
Clearly, also $L_{-1}\tilde\tau_{\alpha+\delta_i-\delta_j}=0$, this gives
(see e.g. [D]):
$$\align
&\sum_{a=1}^s \{
(\delta_a|\alpha+\delta_i-\delta_j)(x_{n_a}^{(a)}-{\delta_{aj}\over
n_jz^{n_j}})+
{1\over 2n_a}\sum_{p=1}^{n_a-1} p(n_a-p)(x_p^{(a)}-{\delta_{aj}\over
pz^p})\times \\
&(x_{n_a-p}^{(a)}-{\delta_{aj}\over (n_j-p)z^{n_j-p}})+{1\over
n_a}\sum_{k=1}^{\infty}(k+n_a)(x_{k+n_a}^{(a)}-{\delta_{aj}\over
(k+n_j)z^{k+n_j}}){\partial\over\partial x_k^{(a)}}\}
\tilde\tau_{\alpha+\delta_i-\delta_j}=0.
\tag{5.2.3}
\endalign$$
So, in a similar way as in [D], one deduces from (5.2.2 and 3) that
$$\tilde\tau_{\alpha+\delta_i-\delta_j}\tau_{\alpha}^{-2}L_{-1}\tau_{\alpha}
-\tau_{\alpha}^{-1}L_{-1}\tilde\tau_{\alpha+\delta_i-\delta_j}=0.
$$
Hence, we find that for all $\alpha\in Q$ and $1\le i,j\le s$:
$$\align
\ &{1\over n_j}
\{ {1\over\tau_{\alpha}}(\sum_{k=1}^{\infty}{\partial\over\partial x_k^{(j)}}
z^{-k-n_j}+(\delta_{ij}-1+(\delta_j|\alpha)+{1\over 2}-{{n_a}\over
2})z^{-n_j}-n_jx^{(i)}_{n_i}
)\tilde{\tau}_{\alpha+\delta_i-\delta_j}\\
&+\sum_{k=1}^{n_j} kx_k^{(j)}{{\tilde{\tau}_{\alpha+\delta_i-\delta_j}}\over
\tau_{\alpha}}z^{k-n_j}
-\sum_{a=1}^s {{n_j}\over n_a}\sum_{k=1}^{\infty} (k+n_a)x_{k+n_a}^{(a)}
{\partial\over \partial
x_k^{(a)}}({{\tilde{\tau}_{\alpha+\delta_i-\delta_j}}\over \tau_{\alpha}})
\}=0.
\tag{5.2.4}
\endalign
$$
Comparing this with (5.1.3), one finds
$$S((\sum_{a=1}^s \{ ({1\over n_a}M(\alpha)L(\alpha)^{1-n_a}C^{(a)}(\alpha))_-
-{{n_a-1}\over 2n_a}L(\alpha)^{-n_a}C^{(a)}(\alpha)\} P(\alpha)R(\alpha))_{ij})
=0
$$
We  thus conclude that the string equation induces for all $\alpha\in Q$:
$$
\sum_{a=1}^s \{ ({1\over n_a}M(\alpha)L(\alpha)^{1-n_a}C^{(a)}(\alpha))_-
-{{n_a-1}\over 2n_a}L(\alpha)^{-n_a}C^{(a)}(\alpha)\}
=0.
\tag{5.2.5}
$$
So, if (5.2.5) holds
$$N(\alpha):=
\sum_{a=1}^s \{ {1\over n_a}M(\alpha)L(\alpha)^{1-n_a}C^{(a)}(\alpha)
-{{n_a-1}\over 2n_a}L(\alpha)^{-n_a}C^{(a)}(\alpha)\}
$$
is a differential operator that satisfies
$$[\sum_{a=1}^s L(\alpha)^{n_a}C^{(a)}(\alpha),N(\alpha)]=1.$$

\vskip 10pt
\subheading{\S 6. $W_{1+\infty}$ constraints}

\vskip 10pt
{\bf 6.1.} Let $e_i$, $1\le i\le s$ be a basis of ${\Bbb C}^s$.
In a similar way as in \S 2, we identify $({\Bbb C}[t,t^{-1}])^s$ with ${\Bbb
C}^{\infty}$,
viz., we put
$$v^{(a)}_{-k-\frac{1}{2}}=t^ke_a.\tag{6.1.1}$$
We can associate to $({\Bbb C}[t,t^{-1}])^s$ $s$--copies of the Lie algebra
of differential operators on ${\Bbb C}^{\times}$,
it has as basis the operators (see [Ra] or [KRa]):
$$-t^{k+\ell}({\partial\over\partial t})^{\ell}e_{ii},
\quad\text{for}\ k\in {\Bbb Z},\ \ell\in{\Bbb Z}_+,\ 1\le i\le s.$$
We will denote this Lie algebra by $D^s$. Via (6.1.1) we can embed this algebra
into $\overline{gl_{\infty}}$
and also into $a_{\infty}$, one finds
$$-t^{k+\ell}({\partial\over\partial t})^{\ell}e_{ii}
\mapsto \sum_{m\in{\Bbb
Z}}-m(m-1)\cdots(m-\ell+1)E^{(ii)}_{-m-k-\frac{1}{2},-m-\frac{1}{2}}.
\tag{6.1.2}
$$
It is straightforward, but rather tedious, to calculate the corresponding
2--cocycle, the result is as follows (see also [Ra] or [KRa]).
Let $f(t), g(t)\in {\Bbb C}[t,t^{-1}]$ then
$$\mu(f(t)({\partial\over\partial t})^{\ell}e_{aa},g(t)({\partial\over\partial
t})^m e_{bb})
=\delta_{ab}{{\ell
!m!}\over(\ell+m+1)!}\text{Res}_{t=0}dt\ f^{(m+1)}(t)g^{(\ell)}(t).
$$
Hence in this way we get a central extension $\hat D^s=D^s\oplus {\Bbb C}c$ of
$D^s$
with Lie bracket
$$\align
&[f(t)({\partial\over\partial t})^{\ell}e_{aa}+\alpha
c,g(t)({\partial\over\partial t})^m e_{bb}+\beta c]=\\
&\delta_{ab}\{
(f(t)({\partial\over\partial t})^{\ell}g(t)({\partial\over\partial t})^m -
g(t)({\partial\over\partial t})^m f(t)({\partial\over\partial t})^{\ell})e_{aa}
+{{\ell !m!}\over(\ell+m+1)!}\text{Res}_{t=0}dt\ f^{(m+1)}(t)g^{(\ell)}(t)c\}.
\tag{6.1.3}\endalign
$$
Since we have the representation $\hat r$ of $a_{\infty}$, we find that
$$\hat r (-t^{k+\ell}({\partial\over\partial t})^{\ell}e_{aa})=
\sum_{m\in{\Bbb Z}}m(m-1)\cdots (m-\ell+1):\psi^{+(a)}_{-m-\frac{1}{2}}
\psi^{-(a)}_{m+k+\frac{1}{2}}:.
$$
In terms of the fermionic fields (2.1.2), we find
$$\sum_{k\in{\Bbb Z}}\hat r (-t^{k+\ell}({\partial\over\partial
t})^{\ell}e_{aa})z^{-k-\ell-1}=
:{{\partial^{\ell}\psi^{+(a)}(z)}\over z^{\ell}}
\psi^{-(a)}(z):. \tag{6.1.4}
$$

\vskip 10pt
{\bf 6.2.} We will now express $-t^{k+\ell}({\partial\over\partial
t})^{\ell}e_{aa}$
in terms of the oscillators $\alpha^{(a)}_k$. For this purpose, we
first calculate
$$\align
:(y-z)\psi^{+(a)}(y)\psi^{-(a)}(z):&=(y-z)\psi^{+(a)}(y)\psi^{-(a)}(z)-1 \\
&=X_a(y,z)-1,\\
\endalign$$
where
$$X_a(y,z)={(\frac{y}{z})}^{\alpha^{(a)}_0}
\exp(-\sum_{k<0}\frac{\alpha^{(a)}_k}{k}(y^{-k}-z^{-k}))
\exp(-\sum_{k>0}\frac{\alpha^{(a)}_k}{k}(y^{-k}-z^{-k})).
\tag{6.2.1}
$$
Then
$$:{{\partial^{\ell}\psi^{+(a)}(z)}\over \partial z^{\ell}}\psi^{-(a)}(z):
=\frac{1}{\ell+1} \frac{\partial^{\ell+1}X_a(y,z)}{\partial
y^{\ell+1}}|_{y=z}.
\tag{6.2.2}
$$
Notice that the right--hand--side of this formula is some normal
ordered expression in the $\alpha^{(a)}_k$'s. For some explicit formulas of
(6.2.2), we refer to the appendix of [AV].

\vskip 10pt
{\bf 6.3.} In the rest of this section, we will show that $\hat D^s$ has a
subalgebra
that will provide the extra constraints, the so called $W$--algebra constraints
on $\tau$.

{\it From now on we assume that $\tau$ is a $\tau$--function of the
$[n_1,n_2,\ldots,n_s]$-th
reduced KP hierarchy, which satisfies the string equation.}
So, we assume that (4.2.3) and (5.2.1) holds. Hence,
for all  $\alpha\in \text{supp}\ \tau$ both
$$\align
Q(\alpha):=&\sum_{a=1}^s L(\alpha)^{n_a}C^{(a)}(\alpha)\quad\text{and} \\
 N(\alpha)=&
\sum_{a=1}^s \{ {1\over n_a}M(\alpha)L(\alpha)^{1-n_a}C^{(a)}(\alpha)
-{{n_a-1}\over 2n_a}L(\alpha)^{-n_a}C^{(a)}(\alpha)\}
\endalign
$$
are differential operators. Thus, also
$N(\alpha)^pQ(\alpha)^q$ is a differential operator, i.e.,
$$((\sum_{a=1}^s \{ {1\over n_a}M(\alpha)L(\alpha)^{1-n_a}
-{{n_a-1}\over 2n_a}L(\alpha)^{-n_a})^pL(\alpha)^{qn_a}C^{(a)}(\alpha))_-=0
\quad\text{for } p,q\in {\Bbb Z}_+.\tag{6.3.1}$$
Using (6.3.1),we are able to prove the following
\proclaim{Lemma 6.1}
For all $\alpha\in Q$ and $p,q\in {\Bbb Z}_+$
$$\text{Res}_{z=0}dz\sum_{a=1}^s z^{qn_a}({1\over n_a}
z^{{1-n_a}\over 2}{\partial\over \partial z}z^{{1-n_a}\over 2})^p
V^+(\alpha,x,z)E_{aa}^{\phantom{aa}t} V^-(\alpha,x',z) =0.
\tag{6.3.2}$$
\endproclaim
\demo{Proof}
Using Taylor's formula we rewrite the right--hand--side of (6.3.2):
$$\text{Res}_{z=0}dz\sum_{a=1}^s z^{qn_a}({1\over n_a}
z^{{1-n_a}\over 2}{\partial\over \partial z}z^{{1-n_a}\over 2})^p
V^+(\alpha,x,z)E_{aa}
\exp (\sum_{\ell =1}^s\sum_{k=1}^{\infty}(x_k^{(\ell)\prime}-x_k^{(\ell)})
{\partial\over\partial x_k^{(\ell)}})^tV^-(\alpha,x,z).
\tag{6.3.3}$$
Since
$${{\partial V^-(\alpha,x,z)}\over\partial x_k^{(\ell)}}=
-(P^-(\alpha,x,z)E_{\ell\ell}\partial^k P^-(\alpha,x,z)^{-1})_+
V^-(\alpha,x,z),
$$
it suffices to prove that for all $m\ge 0$
$$\text{Res}_{z=0}dz\sum_{a=1}^s z^{qn_a}({1\over n_a}
z^{{1-n_a}\over 2}{\partial\over \partial z}z^{{1-n_a}\over 2})^p
V^+(\alpha,x,z)E_{aa}\partial^{mt}V^-(\alpha,x,z) =0.
\tag{6.3.4}$$
Now, let
$\sum_{a=1}^s z^{qn_a}({1\over n_a}
z^{{1-n_a}\over 2}{\partial\over \partial z}z^{{1-n_a}\over 2})^p
V^+(\alpha,x,z)E_{aa}
=\sum_i S_i\partial^{-i}e^{x\cdot z}$
and
\break
$V^-(\alpha,x,z)=\sum_j T_j\partial^{-j}e^{-x\cdot z}$,
then (6.3.4) is equivalent to
$$\align
0&=\text{Res}_{z=0}dz \sum_{i,j} S_i z^{-i}e^{x\cdot z}\partial^m
(e^{-x\cdot zt}T_j(-z)^j)\\
&=\text{Res}_{z=0}dz \sum_{i,j}\sum_{\ell=0}^m (-1)^{m-\ell+j}{m\choose\ell}
S_i\partial^{\ell}(^tT_j)z^{m-i-j-\ell}\\
&=\sum_{\scriptstyle 0\le \ell\le m\atop\scriptstyle i+j+\ell=m+1}
(-1)^{\ell+j}{m\choose \ell}S_i\partial^{\ell}(^tT_j).
\tag{6.3.5}\endalign
$$
On the other hand (6.3.1) implies that
$$\align
0&=
\sum_i S_i\partial^{-i}\sum_j(-\partial)^{-jt}T_j)_-\\
&=\sum_{\scriptstyle i,j\atop\scriptstyle \ell\ge 0}(-1)^j{-i-j\choose \ell}
S_i\partial^{\ell}(^tT_j)\partial^{-i-j-\ell})_-.
\endalign
$$
Now let $i+j+\ell=m+1$, then we obtain that for every $m\ge 0$
$$\align
0&=\sum_{\scriptstyle 0\le \ell\atop\scriptstyle i+j+\ell=m+1}
(-1)^{j}{-i-j\choose \ell}S_i\partial^{\ell}(^tT_j)\\
&=\sum_{\scriptstyle 0\le \ell\le m\atop\scriptstyle i+j+\ell=m+1}
(-1)^{\ell+j}{m\choose \ell}S_i\partial^{\ell}(^tT_j),
\endalign
$$
which proves (6.3.5)\ \ \ $\square$
\enddemo

Taking the $(i,j)$--th coefficient of (6.3.3) one obtains
\proclaim{Corollary 6.2}
For all $\alpha\in Q$, $1\le i,j\le s$  and $p,q\in{\Bbb Z}_+$ one has
$$\text{Res}_{z=0}dz\sum_{a=1}^s z^{qn_a}({1\over n_a}
z^{{1-n_a}\over 2}{\partial\over \partial z}z^{{1-n_a}\over 2})^p
\psi^{+(a)}(z)\tau_{\alpha+\delta_i-\delta_a}\otimes
\psi^{-(a)}(z)\tau_{\alpha+\delta_a-\delta_j}=0.
\tag{6.3.6}
$$
\endproclaim
Notice that (6.3.6) can be rewritten as infinitely many generating series
of Hirota bilinear equations, for the case that $p=q=0$ see [KV].

\vskip 10pt
{\bf 6.4.} The following lemma gives a generalization of an
identity of Date, Jimbo, Kashiwara and Miwa [DJKM3] (see also
 [G]):
\proclaim
{Lemma 6.3}
Let $X_b(y,w)$ be given by (6.2.1), then
$$
\align
\text{Res}_{z=0}dz&\sum_{a=1}^s
\psi^{+(a)}(z)
X_b(y,w)\tau_{\alpha+\delta_i-\delta_a}e^{\alpha+\delta_i-\delta_a}
\otimes\psi^{-(a)}(z)\tau_{\alpha+\delta_a-\delta_j}e^{\alpha+\delta_a-\delta_j}=\\
&(w-y)\psi^{+(b)}(y)\tau_{\alpha+\delta_i-\delta_b}e^{\alpha+\delta_i-\delta_b}
\otimes\psi^{-(b)}(w)\tau_{\alpha+\delta_b-\delta_j}e^{\alpha+\delta_b-\delta_j}.
\tag{6.4.1}\endalign
$$
\endproclaim
\demo
{Proof}
The left--hand--side of (6.4.1) is equal to
$$\align\text{Res}_{z=0}&dz
\sum_{a=1}^s \epsilon(\delta_a, \delta_i+\delta_j)
z^{(\delta_a|\delta_i+\delta_j)-2}
y^{(\delta_b|\alpha+\delta_i-\delta_a)}
w^{-(\delta_b|\alpha+\delta_i-\delta_a)}
({{z-y}\over z-w})^{\delta_{ab}}\times\\
&e^{x^{(b)}\cdot y- x^{(b)}\cdot w+x^{(a)}\cdot z}
\exp (-\sum_{k=1}^{\infty}{1\over k}(y^{-k}-w^{-k})
{\partial\over\partial x^{(b)}_k}+{1\over k}z^{-k}
{\partial\over\partial x^{(a)}_k})
\tau_{\alpha+\delta_i-\delta_a}e^{\alpha+\delta_i}
\otimes \\
&e^{-x^{(a)}\cdot z}
\exp (\sum_{k=1}^{\infty}{1\over k}z^{-k}
{\partial\over\partial x_k^{(a)}})
\tau_{\alpha+\delta_a-\delta_j}e^{\alpha-\delta_j}.
\endalign
$$

Recall the bilinear identity for $\beta=\alpha$:
$$\text{Res}_{z=0}dz\psi^{+(a)}(z)
\tau_{\alpha+\delta_i-\delta_a}e^{\alpha+\delta_i-\delta_a}
\otimes
\psi^{-(a)}(z)\tau_{\alpha+\delta_a-\delta_j}e^{\alpha+\delta_a-\delta_j}
=0.
$$
Let $X_b(y,w)\otimes 1$ act on this identity, then
$$\align
\text{Res}_{z=0}dz&
\sum_{a=1}^s \epsilon(\delta_a, \delta_i+\delta_j)
z^{(\delta_a|\delta_i+\delta_j)-2}
y^{(\delta_b|\alpha+\delta_i)}
w^{-(\delta_b|\alpha+\delta_i)}
({{y-z}\over w-z})^{\delta_{ab}}
e^{x^{(b)}\cdot y- x^{(b)}\cdot w+x^{(a)}\cdot z}\times \\
&\exp (-\sum_{k=1}^{\infty}{1\over k}(y^{-k}-w^{-k})
{\partial\over\partial x^{(b)}_k}+{1\over k}z^{-k}
{\partial\over\partial x^{(a)}_k})
\tau_{\alpha+\delta_i-\delta_a}e^{\alpha+\delta_i}
\otimes \\
&e^{-x^{(a)}\cdot z}
\exp (\sum_{k=1}^{\infty}{1\over k}z^{-k}
{\partial\over\partial x_k^{(a)}})
\tau_{\alpha+\delta_a-\delta_j}e^{\alpha-\delta_j}
=0.
\endalign$$
Now, using this and the fact that
${{z-y}\over z-w}={{1-y/z}\over1-w/z}={{w-y}\over z}\delta(w/z) +
({y\over w}){{1-z/y}\over 1-z/w}$,
where
$\delta(w/z)=\sum_{k\in{\Bbb Z}}(w/z)^k$
(such that $f(w,z)\delta(w/z)=f(w,w)\delta(w/z)$),
we obtain that the left--hand--side  of (6.4.1) is equal to
$$\align
(w-y)&
y^{(\delta_b|\alpha+\delta_i-\delta_b)}
e^{x^{(b)}\cdot y}
\exp (-\sum_{k=1}^{\infty}{1\over k}y^{-k}
{\partial\over\partial x^{(b)}_k}
\tau_{\alpha+\delta_i-\delta_b}e^{\alpha+\delta_i}
\otimes
w^{-(\delta_b|\alpha+\delta_b-\delta_j)}
e^{-x^{(b)}\cdot w}\times \\
&\exp (\sum_{k=1}^{\infty}{1\over k}w^{-k}
{\partial\over\partial x_k^{(b)}})
\tau_{\alpha+\delta_b-\delta_j}e^{\alpha-\delta_j},
\endalign
$$
which is equal to the right--hand--side of (6.4.1)\ \ \ $\square$
\enddemo

Define $c_b(\ell,p)$ as follows
$$(z^{{1-n_b}\over 2}{\partial\over\partial z}z^{{1-n_b}\over 2})^p
=\sum_{\ell=0}^p c_b(\ell,p)z^{-n_bp+\ell}({\partial\over\partial z})^{\ell}.
\tag{6.4.2}
$$
One also has
$$(M(\alpha)L(\alpha)^{-n_b +1}-{{n_b-1}\over 2}L(\alpha)^{-n_b})^p=
\sum_{\ell=0}^p c_b(\ell,p) M(\alpha)^{\ell}L(\alpha)^{-n_bp+\ell}.
\tag{6.4.3}
$$
Then it is straightforward to show that
$$\align
c_b(\ell,p)=&\sum_{0\le q_0<q_1\cdots<q_{p-\ell-1}\le p-1}
[(q_0+\frac{1}{2})(1-n_b)]\times \\
&[(q_1+\frac{1}{2})(1-n_b)-1]\cdots
[(q_{p-\ell -1}+\frac{1}{2})(1-n_b)-(p-\ell-1)].
\tag{6.4.4}\endalign
$$
Now using (6.4.2) and removing the tensor product symbol in
(6.3.6), where we write $x$, respectively $x'$, for the first,
respectively the second, component of the tensor product, one
gets:
$$\text{Res}_{z=0}dz \sum_{a=1}^s (\frac{1}{n_a})^pz^{qn_a}
\sum_{\ell=0}^p c_a(\ell,p) z^{-n_ap+\ell}
\frac{\partial^{\ell}\psi^{+(a)}(z)}{\partial z^{\ell}}
\tau_{\alpha+\delta_i-\delta_a}(x)
\psi^{-(a)}(z)'\tau_{\alpha+\delta_a-\delta_j}(x')
=0
$$
Using Lemma 6.3, this is equivalent to
$$\align
\text{Res}_{\scriptstyle y=0\atop\scriptstyle z=0}dydz&
\sum_{a=1}^s \psi^{+(a)}(y)
\sum_{b=1}^s
(\frac{1}{n_b})^pz^{qn_b}
\sum_{\ell=0}^p \frac{c_b(\ell,p)}{\ell+1} z^{-n_bp+\ell}
\frac{\partial^{\ell+1}X_b(w,z)}{\partial z^{\ell+1}}|_{w=z}\times \\
&\tau_{\alpha+\delta_i-\delta_a}(x)e^{\alpha+\delta_i-\delta_a}
\psi^{-(a)}(y)'\tau_{\alpha+\delta_a-\delta_j}(x')(e^{\alpha+\delta_a-\delta_j})'
=0.
\endalign
$$
Now, recall (6.1.4) and (6.2.2), then
$$\align
\ \ \ &\text{Res}_{z=0}dz
\sum_{b=1}^s
(\frac{1}{n_b})^pz^{qn_b}
\sum_{\ell=0}^p \frac{c_b(\ell,p)}{\ell+1} z^{-n_bp+\ell}
\frac{\partial^{\ell+1}X_b(w,z)}{\partial w^{\ell+1}}|_{w=z}
=\\
&\text{Res}_{z=0}dz
\sum_{b=1}^s
(\frac{1}{n_b})^pz^{qn_b}
\sum_{\ell=0}^p c_b(\ell,p) z^{-n_bp+\ell}
:\frac{\partial^{\ell}\psi^{+(b)}(z)}{\partial z^{\ell}}\psi^{-(b)}(z):
=\\
&\sum_{b=1}^s
(\frac{1}{n_b})^p
\sum_{\ell=0}^p {c_b(\ell,p)}
\hat r (-t^{(q-p)n_b+\ell}(\frac{\partial}{\partial t})^{\ell}e_{bb})
=\\
&-\sum_{b=1}^s
(\frac{1}{n_b})^p
\hat r (t^{n_bq}(t^{\frac{1-n_b}{2}}\frac{\partial}{\partial t}
t^{\frac{1-n_b}{2}})^pe_{bb})
=\\
&-\sum_{b=1}^s
\hat r ((t^{\frac{n_b-1}{2}}t^{n_bq}(\frac{\partial}{\partial t^{n_b}})^p
t^{\frac{1-n_b}{2}})e_{bb})
=\\
&\sum_{b=1}^s
\hat r (t^{\frac{n_b-1}{2}}(-\lambda_b^{q}(\frac{\partial}{\partial
\lambda_b})^p)
t^{\frac{1-n_b}{2}})e_{bb})\qquad\text{where}\ \ \ \ \lambda_b=t^{n_b}
\\
\overset{def}\to{=}&W^{(p+1)}_{q-p}.\tag{6.4.5}
\endalign
$$
Hence, (6.3.6) is equivalent to
$$\text{Res}_{y=0}dy
\sum_{a=1}^s \psi^{+(a)}(y)
W^{(p+1)}_{q-p}
\tau_{\alpha+\delta_i-\delta_a}(x)e^{\alpha+\delta_i-\delta_a}
\psi^{-(a)}(y)'\tau_{\alpha+\delta_a-\delta_j}(x')(e^{\alpha+\delta_a-\delta_j})'
=0.
\tag{6.4.6}
$$

If we ignore the cocycle term for a moment, then it is obvious from,
the sixth line of (6.4.5), that the elements $W^{(p+1)}_q$ are the generators
of the W--algebra $W_{1+\infty}$ (the cocycle term, however, will be slightly
different). Upto some  modification of the elements $W^{(p+1)}_0$, one gets
the standard commutation relations of $W_{1+\infty}$, where $c=nI$.

As the next step, we take in (6.4.6) $x_k^{(i)}=x_k^{(i)\prime}$, for all
$k\in{\Bbb N},\ 1\le i\le s$, we then obtain
$$\cases
\frac{\partial}{\partial
x_1^{(i)}}(\frac{W^{(p+1)}_{q-p}\tau_{\alpha}}{\tau_{\alpha}})=0&
\text{if } i=j,\\
\tau_{\alpha+\delta_i-\delta_j}W^{(p+1)}_{q-p}\tau_{\alpha}=
\tau_{\alpha}W^{(p+1)}_{q-p}\tau_{\alpha+\delta_i-\delta_j}&
\text{if }i\ne j.
\endcases
\tag{6.4.7}
$$
The last equation means that for all $\alpha,\beta\in \text{supp}\ \tau$ one
has
$$\frac{W^{(p+1)}_{q-p}\tau_{\alpha}}{\tau_{\alpha}}=
\frac{W^{(p+1)}_{q-p}\tau_{\beta}}{\tau_{\beta}}.
\tag{6.4.8}
$$
Next we divide (6.4.6) by $\tau_{\alpha}(x)\tau_{\alpha}(x')$, of course
only for $\alpha\in \text{supp}\ \tau$, and use (6.4.8). Then for all
$\alpha, \beta\in\text{supp}\ \tau$ and $p,q\in {\Bbb Z}_+$ one has
$$\align
\text{Res}_{z=0}dz&
\sum_{a=1}^s \exp(-\sum_{k=1}^{\infty}\frac{z^{-k}}{k}
\frac{\partial}{\partial x_k^{(a)}})
(\frac{W^{(p+1)}_{q-p}\tau_{\beta}(x)}{\tau_{\beta}(x)})
\frac{\psi^{+(a)}(z)\tau_{\alpha+\delta_i-\delta_a}(x)}{\tau_{\alpha}(x)}
e^{\alpha+\delta_i-\delta_a}\times
\\
&\frac{\psi^{-(a)}(z)'\tau_{\alpha+\delta_a-\delta_j}(x')}{\tau_{\alpha}(x')}
(e^{\alpha+\delta_a-\delta_j})'
=0.\endalign
$$
Since one also has the bilinear identity (3.3.3) (see also (2.4.1-2), we can
subtract that part and thus
obtain the following
\proclaim
{Lemma 6.4}
For all
$\alpha, \beta\in\text{supp}\ \tau$ and $p,q\in {\Bbb Z}_+$ one has
$$\align
\text{Res}_{z=0}dz&
\sum_{a=1}^s \{\exp(-\sum_{k=1}^{\infty}\frac{z^{-k}}{k}
\frac{\partial}{\partial x_k^{(a)}})-1\}
(\frac{W^{(p+1)}_{q-p}\tau_{\beta}(x)}{\tau_{\beta}(x)})\times
\\
&\frac{\psi^{+(a)}(z)\tau_{\alpha+\delta_i-\delta_a}(x)}{\tau_{\alpha}(x)}
e^{\alpha+\delta_i-\delta_a}
\frac{\psi^{-(a)}(z)'\tau_{\alpha+\delta_a-\delta_j}(x')}{\tau_{\alpha}(x')}
(e^{\alpha+\delta_a-\delta_j})'
=0.
\tag{6.4.9}\endalign
$$
\endproclaim
Define
$$S(\beta,p,q,x,z):=\sum_{a=1}^s
\{\exp(-\sum_{k=1}^{\infty}\frac{z^{-k}}{k}
\frac{\partial}{\partial x_k^{(a)}})-1\}
(\frac{W^{(p+1)}_{q-p}\tau_{\beta}(x)}{\tau_{\beta}(x)})E_{aa}.
$$
Notice that the first equation of (6.4.7) implies that
$\partial\circ S(\beta,p,q,x,\partial)= S(\beta,p,q,x,\partial)\circ\partial$.
Then viewing (6.4.9) as the $(i,j)$--th entry of a matrix, (6.4.9) is
equivalent to
$$\text{Res}_{z=0}dz P^+(\alpha)R^+(\alpha)S(\beta,p,q,x,\partial)e^{x\cdot z}
\ ^t(P^-(\alpha)'R^-(\alpha)'e^{-x'\cdot z})=0.
\tag{6.4.10}
$$
Now using Lemma 3.1, one deduces
$$(P^+(\alpha)R^+(\alpha)S(\beta,p,q,x,\partial)
R^+(\alpha)^{-1}P^+(\alpha)^{-1})_-=0,
\tag{6.4.11}
$$
hence
$$P^+(\alpha)S(\beta,p,q,x,\partial)
P^+(\alpha)^{-1}=(P^+(\alpha)S(\beta,p,q,x,\partial)
P^+(\alpha)^{-1})_-=0.
$$
So $S(\beta,p,q,x,\partial)=0$
and therefore
$$\{\exp(-\sum_{k=1}^{\infty}\frac{z^{-k}}{k}
\frac{\partial}{\partial x_k^{(a)}})-1\}
(\frac{W^{(p+1)}_{q-p}\tau_{\beta}(x)}{\tau_{\beta}(x)})=0.
$$
{}From which we conclude that
$$W_k^{(p+1)}\tau_{\beta}=\text{constant }\tau_{\beta}\quad\text{for all}\
-k\le p\ge 0.
\tag{6.4.12}
$$
In order to determine the constants on the right--hand--side of (6.4.12)
we calculate the Lie brackets
$$[W_{-1}^{(2)},\frac{-1}{k+p+1}W_{k+1}^{(p+1)}]\tau_{\beta}=0
\tag{6.4.13}
$$
and thus obtain the main result
\proclaim
{Theorem 6.5}
The following two conditions for $\tau\in F^{(0)}$ are equivalent:

\noindent (1) $\tau$ is a $\tau$--function of the $[n_1,n_2,\ldots,n_s]$--th
reduced
$s$--component KP hierarchy which satisfies the string equation (5.2.1).

\noindent (2) For all $-k\le p\ge 0$:
$$(W_k^{(p+1)}+\delta_{k0}c_p)\tau=0,
\tag{6.4.14}$$
where
$$\align
c_p=&\frac{1}{2p+2}
\sum_{a=1}^s (\frac{-1}{n_a})^{p+1}\sum_{\ell=0}^p
\ell\cdot\ell !{n_a+\ell\choose\ell+2}
\sum_{0\le q_0<q_1\cdots<q_{p-\ell-1}\le p-1}
[(q_0+\frac{1}{2})(n_a-1)]\times \\
&[(q_1+\frac{1}{2})(n_a-1)+1]\cdots
[(q_{p-\ell -1}+\frac{1}{2})(n_a-1)+p-\ell-1].
\tag{6.4.15}\endalign
$$
\endproclaim

For $p=0,1$, the constants $c_p$ are equal to 0, respectively
$\sum_{a=1}^s\frac{n_a^2-1}{24n_a}$.

\demo{Proof of Theorem 6.5}
The case (2) $\Rightarrow$ (1) is trivial. For the implication
(1) $\Rightarrow$ (2), we only have to calculate the left--hand--side of
(6.4.13). It is obvious that this is equal to
$(W_k^{(p+1)}+c_{p,k})\tau_{\beta}$, where
$c_{p,k}=\mu(W_{-1}^{(2)},\frac{-1}{k+p+1}W_{k+1}^{(p+1)})$.
It is clear from (6.1.3) that $c_{p,k}=0$ for $k\ne 0$. So from now
on we assume that $k=0$ and $c_p=c_{p,0}$. Then
$$\align
c_p&=\frac{-1}{p+1}\mu(W_{-1}^{(2)},W_1^{(p+1)})
\\
&=\frac{-1}{p+1}\sum_{a=1}^s (\frac{1}{n_a})^{p+1}
\mu(\frac{1-n_a}{2}t^{-n_a}+t^{1-n_a}\frac{\partial}{\partial t},
\sum_{\ell=0}^p c_a(\ell,p)t^{n_a+\ell}(\frac{\partial}{\partial t})^{\ell})
\\
&=\frac{1}{2p+2}
\sum_{a=1}^s (\frac{1}{n_a})^{p+1}
\sum_{\ell=0}^p  (-1)^{\ell+1}
\ell\cdot\ell !{n_a+\ell\choose\ell+2}
c_a(\ell,p),
\endalign
$$
which equals (6.4.15).\ \ \ $\square$
\enddemo

\vskip 10pt

\subheading{\S 7. A geometrical interpretation of the string equation on the
Sato
Grassmannian}

\vskip 10pt
{\bf 7.1.} It is well--known that every $\tau$--function of the
1--component KP hierarchy corresponds to a point of the Sato
 Grassmannian $Gr$ ( see e.g. [S]).
Let $H$ be the space of formal Laurent series $\sum a_n t^n$ such that
$a_n=0$ for $n>>0$. The points of $Gr$ are those linear subspaces $V\subset H$
for which the naturel projection $\pi_+$ of $V$ into
$H_+=\{\sum a_nt^n\in H|a_n=0 \text{ for all } n<0\}$ is a Fredholm
operator. The big cell $Gr^0$ of $Gr$ consists of those $V$ for which $\pi_+$
is an isomorphism.

The connection between $Gr$ and the semi--infinite wedge space is made as
follows.
Identify $v_{-k-\frac{1}{2}}=t^k$.
Let $V$ be a point of $Gr$ and $w_0(t),w_{-1}(t),\ldots $ be a basis of $V$,
then we associate to $V$ the following element in the semi--infinite wedge
space
$$w_0(t)\wedge w_{-1}(t)\wedge w_{-2}(t)\wedge \ldots .$$
If $\tau$ is a $\tau$-- function of the $n$--th KdV hierarchy, then $\tau $
corresponds to a point of $Gr$ that satisfies
$t^nV\subset V$ (see e.g. [SW], [KS]).

In the case of the $s$--component KP hierarchy and its
$[n_1,n_2,\ldots,n_s]$--reduction
we find it convenient to represent the Sato Grassmannian slightly different.
Let now $H$ be the space of formal laurent series $\sum a_nt^n$ such that
$a_n \in {\Bbb C}^s$ and $a_n=0$ for $n>>0$. The points $Gr$ are those linear
subspaces
$V\subset H$ for which the projection $\pi_+$ of $V$ into
$H_+=\{\sum a_nt^n\in H|a_n=0 \text{ for all } n<0\}$ is a Fredholm
operator. Again, the big cell $Gr^0$ of $Gr$ consists of those $V$ for which
$\pi_+$
is an isomorphism. The connection with the semi--infinite wedge space is
of course given in a similar way via (2.1.1):
$$v_{nj-N_a-p+\frac{1}{2}}=v^{(a)}_{n_aj-p+\frac{1}{2}}=t^{-n_aj+p-1}e_a,$$
here $e_a$, $1\le a\le s$ is an orthonormal basis of ${\Bbb C}^s$.

It is obvious that $\tau$--functions of the $[n_1,n_2,\ldots,n_s]$--th reduced
$s$--component KP
hierarchy correspond to those subspaces $V$ for which
$$(\sum_{a=1}^s t^{n_a}E_{aa})V\subset V.
\tag{7.1.1}
$$

\vskip 10pt
{\bf 7.2.} The proof that there exists a $\tau$--function of the
$[n_1,n_2,\ldots n_s]$--th
reduced KP hierarchy that satisfies the string equation is in great details
similar
to the proof of Kac and Schwarz [KS] in the principal case, i.e., the
$n$--th KdV case.

Recall the string equation $L_{-1}\tau=H_{-1}\tau=0$. Now modify the origin
by replacing $x_{n_a+1}$ by $x_{n_a+1}+1$ for all $1\le a\le s$.
Then the string equation transforms to
$$(L_{-1}+\sum_{a=1}^s \frac{n_a+1}{n_a} x_1^{(a)})\tau =0,
$$
or equivalently
$$(H_{-1}+\sum_{a=1}^s \frac{n_a+1}{n_a} x_1^{(a)})\tau =0.
$$
In terms of elements of $\hat D$ this is
$$\hat r(-A)\tau=0\tag{7.2.1}$$
where
$$A=\sum_{a=1}^s\frac{1}{n_a}((n_a+1)t+t^{1-n_a}\frac{\partial}{\partial t}
-\frac{n_a-1}{2}t^{-n_a})E_{aa}.
\tag{7.2.2}
$$
Hence for $V\in Gr$, this corresponds to
$$AV\subset V.\tag{7.2.3}
$$
Now we will prove that there exists a subspace $V$ satisfying
(7.1.1) and (7.2.3).
We will first start by assuming that $m=n_1=n_2=\cdots =n_s$
(this is the case that $L(\alpha)^m$ is a differential operator).
For this case we will show that there exists a unique point in the
big cell $Gr^0$ that satisfies both
(7.1.1) and (7.2.3). So assume that $V\in Gr^0$ and that $V$ satisfies
these two conditions. Since the projection $\pi_+$ on $H_+$ is an isomorphism,
there
exist $\phi_a\in V$, $1\le a\le s$, of the form
$\phi_a=e_a+\sum_{i,a}c_{i,a}t^{-i}$,
with $c_{i,a}=\sum_{b=1}^s c_{i,a}^{(b)} e_b\in {\Bbb C}^s$.
Now $A^p\phi_a=t^pe_a+$lower degree terms, hence these functions for $p\ge 0$
and $1\le a\le s$ form a basis of $V$. Therefore, $t^m\phi_a$ is a linear
combination of
$A^p\phi_b$; it is easy to observe that $A^m\phi_a=\text{constant }t^m\phi_a$.
Using this we find a recurrent relation for the $c_{i,a}^{(b)}$'s:
$$(\frac{m+1}{m})^{m-1}ic_{i,a}^{(b)}
=\sum_{\ell=1}^{m-1} d_{m,i,\ell}c_{i-\ell(m+1),a},
\tag{7.2.4}
$$
here the $d_{m,i,\ell}$ are coefficients depending on $m,i,\ell$,
which can be calculated explicitly using (7.2.2).
Since $c^{(b)}_{0,a}=\delta_{ab}$ and $c^{(b)}_{i,a}=0$ for $i<0$
one deduces from (7.2.4) that $c_{i,a}^{(b)}=0$ if $b\ne a$,  and
$c_{i,a}^{(a)}=0$ if $i\ne (m+1)k$ with $k\in{\Bbb Z}$.
So the $\phi_a$ for $1\le a\le s$ can be determined uniquely.
More explicitly, all $\phi_a$ are of the form $\phi_a=\phi^{(m)}e_a$,
with
$$\phi^{(m)}=\sum_{i=1}^{\infty} b_i^{(m)}t^{-(m+1)i},
\tag{7.2.5}
$$
where the $b_i$ do not depend on $a$ and satisfy
$$(\frac{m+1}{m})^{m-1} i(m+1)b_i^{(m)}=\sum_{\ell=1}^{m-1} d_{m,i, \ell}
b^{(m)}_{i-\ell}.
$$
Thus the space $V\in Gr^0$ is spanned by $t^{km}A^{\ell}\phi_a$ with
$1\le a\le s$ , $k\in {\Bbb Z}_+$, $0\le \ell<m$.

Notice  that in the case that all $n_a=1$ we find that $V=H_+$,
meaning that the only solution of
(7.1.1) and (7.2.3) in $Gr^0$ is $\tau=\text{constant }e^0$,
corresponding to the vacuum vector $|0\rangle$.

If not all $n_a$ are the same, then it is ovious that there still is a $V\in
Gr^0$
satisfying (7.1.1) and (7.2.3), viz., $V$ spanned by
$t^{kn_a}A^{\ell_a}\phi^{(n_a)}e_a$,
with $1\le a\le s$, $k\in{\Bbb Z}_+$, $0\le \ell_a<n_a$,
where $\phi^{(n_a)}$ is the unique solution determined by (7.2.5).
However, at the present moment we do not know if this $V\in Gr^0$ is still
unique in $Gr^0$.

\break

\enddocument